\documentclass[11pt]{article}
\usepackage{moriond,epsfig}

\def\NIMA{{\em Nucl. Instrum. Methods} A }

\def\PLB{{\em Phys. Lett.}  B }
\def\PRL{\em Phys. Rev. Lett. }
\def\PRD{{\em Phys. Rev.} D }

\def\be{\begin{equation}}
\def\ee{\end{equation}}
\def\bea{\begin{eqnarray}}
\def\eea{\end{eqnarray}}
\def\fbm{fb$^{-1}$}

\def\BF{\ensuremath{{\mathcal B}}}
\def\LK{\ensuremath{{\cal L}}}
\def\RL{\ensuremath{\cal R}}
\def\ks{\ensuremath{K_S^0}}
\def\lcppi{\ensuremath{B^-\to \Lambda_c^+ \bar{p} \pi^-}}
\def\llk{\ensuremath{B^+\to \Lambda \bar{\Lambda} K^+}}
\def\llpi{\ensuremath{B^+\to \Lambda \bar{\Lambda} \pi^+}}
\def\ppk{\ensuremath{B\to p \bar{p} K}}
\def\ppkp{\ensuremath{B^+\to p \bar{p} K^+}}
\def\ppks{\ensuremath{B^+\to p \bar{p} K_S}}
\def\sppk{\ensuremath{p \bar{p} K}}
\def\plpi{\ensuremath{B^0\to p \bar{\Lambda} \pi^-}}
\def\plg{\ensuremath{B^+\to p \bar{\Lambda} \gamma}}
\def\de{\ensuremath{\Delta E}}
\def\mb{\ensuremath{M_{\mbox{\scriptsize bc}}}}
\begin{document}
\vspace*{4cm}
\title{BARYONIC $B$ DECAYS}

\author{ J. SCH\"UMANN, for the Belle Collaboration}

\address{Department of Physics, High Energy Physics Group, National Taiwan
University \\
No.1, Sec.4 Roosevelt Rd., 106 Taipei, Taiwan}

\maketitle\abstracts{
We summarize recent results of baryonic $B$ decays from Belle and BaBar.
The results from Belle are based on 140 \fbm{} and results from BaBar are based
on 81 \fbm {} of data collected at the $\Upsilon(4S)$ resonance at KEKB or PEPII
respectively. We report the results of two- and three-body baryonic $B$ decays
as well as searches for pentaquarks. The three-body baryonic $B$ decays display
an enhancement in the low mass region, which is not in agreement with general 
phase space expectations. 
}

\section{Introduction}
Baryonic $B$ decays are a unique feature of $B$ meson decays and have already
been well established through previous measurements~\cite{ppk,plpi,pph,LLK}. 
Three-body baryonic $B$ decays were found to have a larger branching fraction 
than two-body decays. 

Three-body baryonic $B$ decays display the common feature of a peaking behaviour
toward the baryon-antibaryon pair mass threshold. This feature has also been
observed in baryonic $J/\Psi$ decays~\cite{bib:Jpsi}, 
indicating it may be a universal feature of these decays. Possible
explanations include intermediate (gluonic) resonant states or non-perturbative
QCD effects of the quark fragmentation
process~\cite{HS,rhopn,glueball,RosnerB}. Angular
distributions are used to discriminate between the different decay mechanisms.

\section{The analyses}
\subsection{Event Selection}
We use a data sample of $152\times 10^6$ $B\overline{B}$ pairs,
corresponding to an integrated luminosity of 140
fb$^{-1}$, collected by 
the Belle detector 
at the KEKB~\cite{KEKB} asymmetric energy $e^+e^-$ 
collider. The detector is described in detail
elsewhere~\cite{Belle}.

We select events through the following decay channels: $\Lambda_c^+ \to p K^-
\pi^+$, $p \bar{K}^0$, $\Lambda \pi^+$, $p \bar{K}^- \pi^+\pi^-$ and $\Lambda
\pi^+ \pi^- pi^+$; $\Lambda \to p \pi^-$ and $K_S^0 \to \pi^+
\pi^-$.\footnote{Charge conjugate modes are implicitly included throughout this
paper}. 

All primary charged tracks are required to satisfy track quality criteria based
on the track impact parameters relative to the interaction point (IP). 
To identify charged
tracks, the proton ($L_p$), kaon ($L_K$) and pion ($L_{\pi}$) likelihoods are
determined from information obtained by the hadron identification system. A
track is identified as a proton if $L_p / (L_p + L_{\pi}) > 0.6$ and $ L_p /
(L_p + L_K) > 0.6$ ($0.01$), or as a kaon if $L_K / (L_K + L_{\pi}) >0.6$ 
($0.2$), or as a pion if $L_{\pi} / (L_K + L_{\pi}) >0.6$ ($0.05$).

We require the mass of the $\Lambda$ candidates to be consistent with the 
nominal $\Lambda$ baryon mass, $1.111$ $(1.113)$ GeV $< M_{p\pi^-} < $ $1.121$
$(1.118)$ GeV. 
The values inside (without) brackets apply to the $B\to \Lambda_c^+ \bar{p}
\pi^-$ (all other) analyses.

Continuum background, the major background contribution for all decays, is
suppressed in the \lcppi{} decay by imposing requirements on the angle between
the thrust axis of the $B$ candidate tracks and that of other tracks, as well as
on the ratio of the second to the zeroth Fox-Wolfram moment~\cite{Fox}. For
all other decays, we form a Fisher discriminant\cite{fisher} that combines
seven event shape variables. Probability density
functions (PDF) for the Fisher discriminant and the cosine of the 
angle between the
$B$ flight direction and the beam direction in the $\Upsilon(4S)$ rest frame are
combined to form the signal (background) likelihood $\LK_s$ ($\LK_b$). We then
optimize the selection of the likelihood ratio $\RL = \LK_s / (\LK_s + \LK_b)$ 
for each mode. For more details of the selection and background suppression
refer to~\cite{bib:lamc,bib:pp,LLK,bib:lowm,bib:plg}.

\subsection{Charmless $B$ decays}

We perform an unbinned likelihood fit that maximizes the likelihood function, 
$$ L = {e^{-(N_s+N_b)} \over N!}\prod_{i=1}^{N} 
\left[\mathstrut^{\mathstrut}_{\mathstrut}N_s P_s(M_{{\rm bc}_i},\Delta{E}_i)+
N_b P_b(M_{{\rm bc}_i},\Delta{E}_i)\right],$$
to estimate the signal yield in 
5.20 GeV/$c^2 < \mb < 5.29$ GeV/$c^2$ and $-0.1$ GeV $ < \de< 0.2$
GeV for \ppk{} and \plpi, $-0.15$ GeV $ < \de< 0.3$ for \llk{} and 
$-0.2$ GeV $ < \de< 0.5$ for \plg;
here $P_s\ (P_b)$ denotes the signal (background) PDF, 
$N$ is the number of events in the fit, and $N_s$ and $N_b$
are fit parameters representing the number of signal and background
events, respectively.

\begin{figure}
\vskip 0.2cm
\psfig{figure=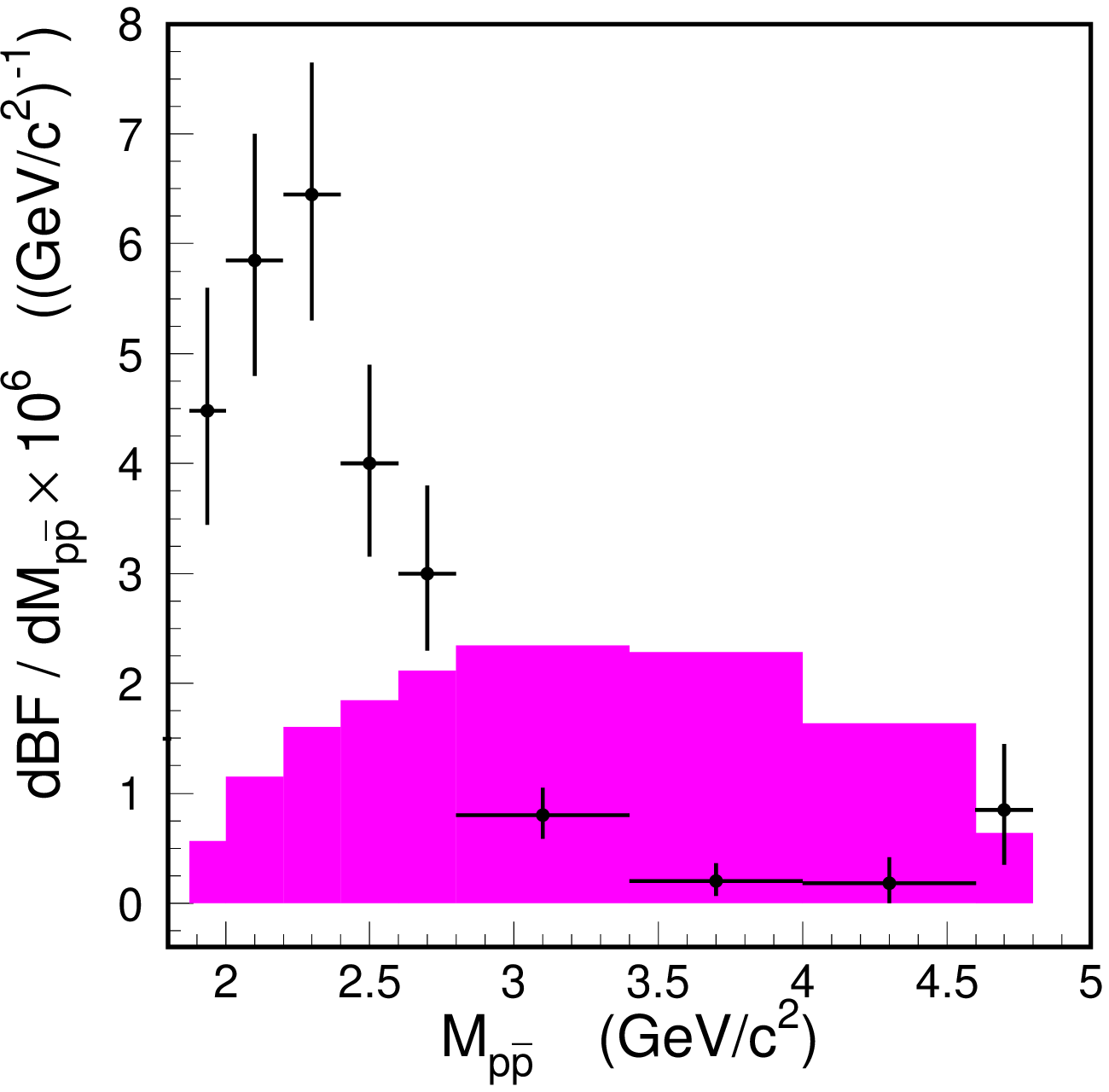,height=1.2in}
\psfig{figure=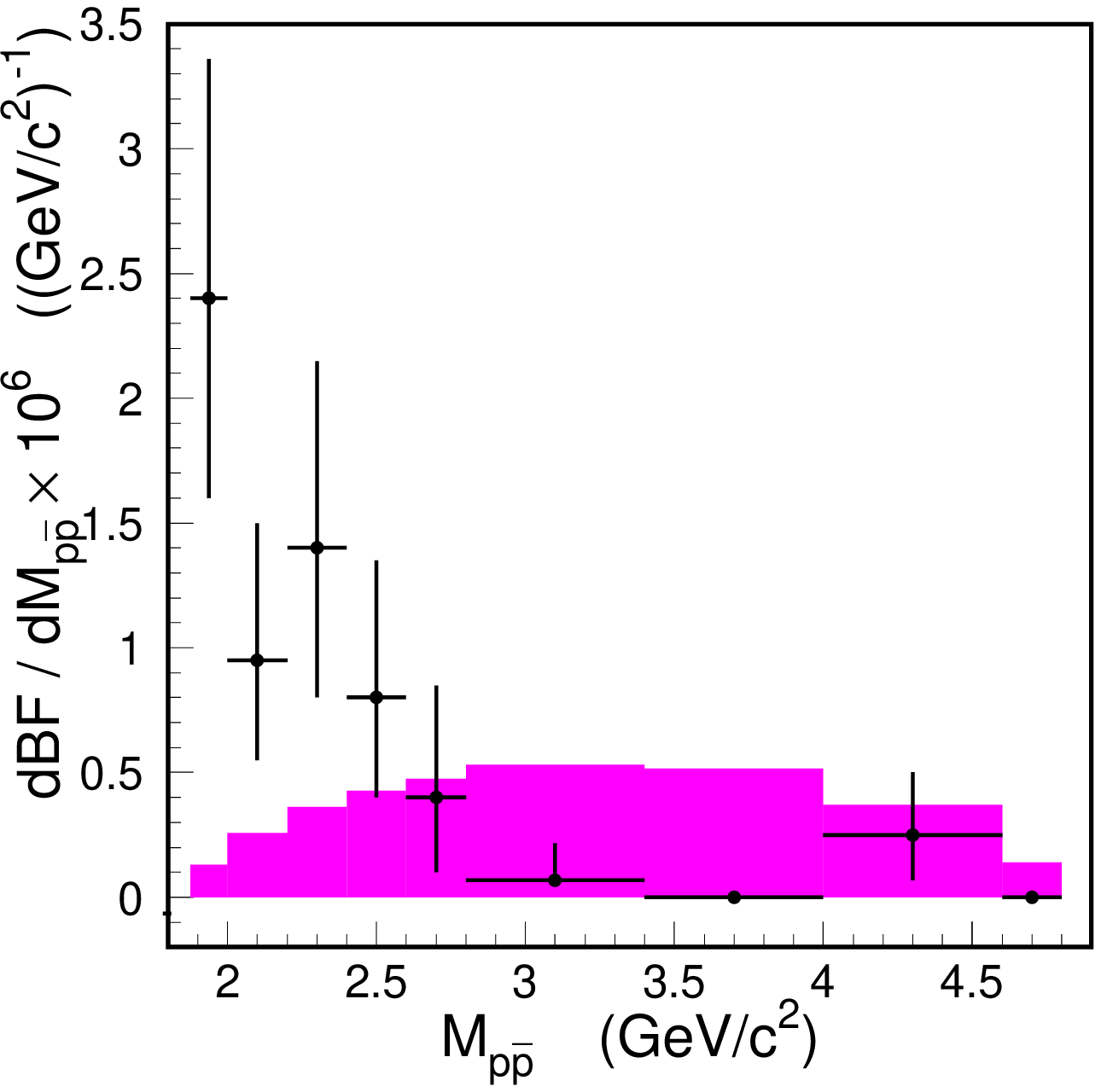,height=1.2in}
\psfig{figure=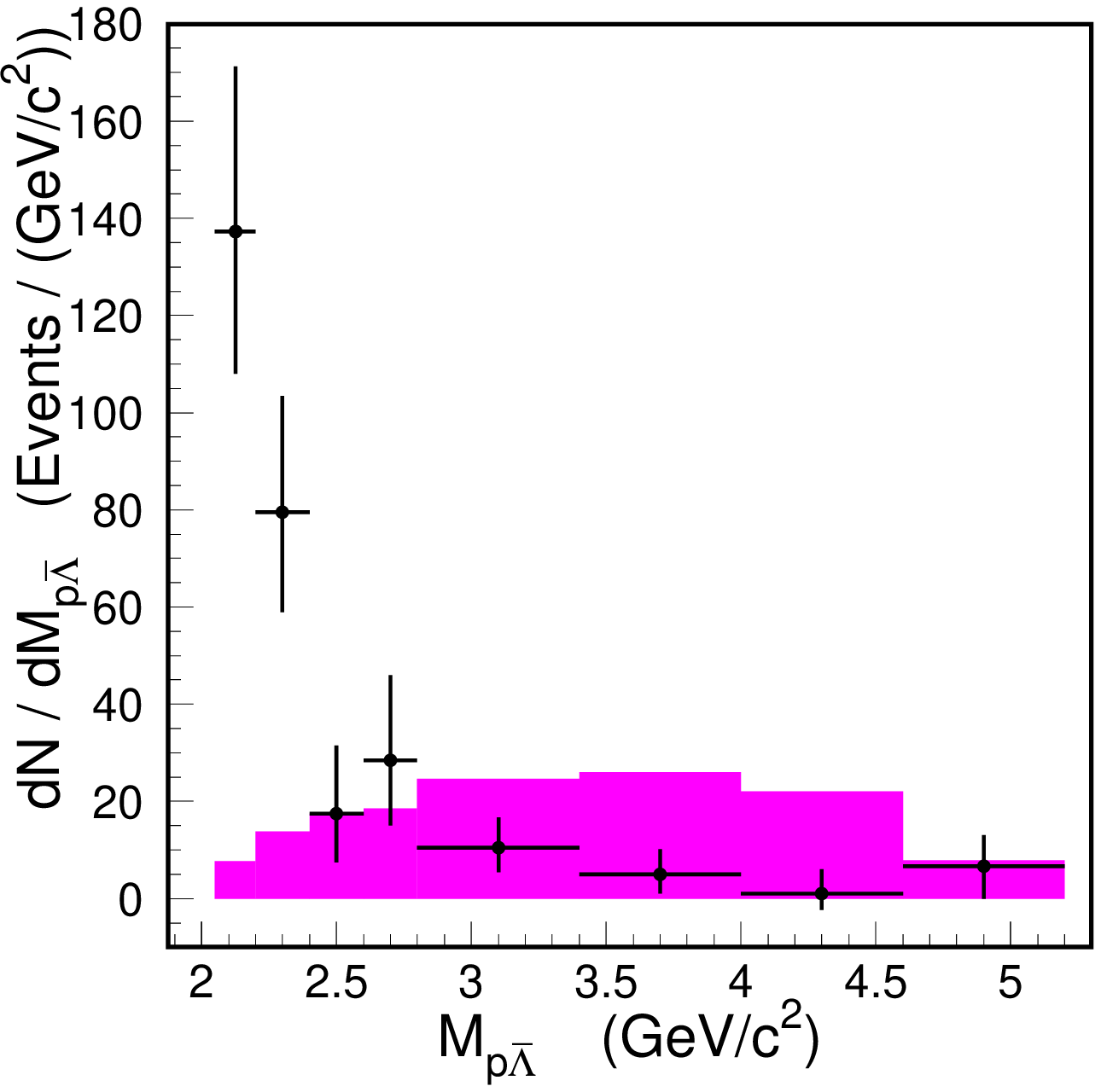,height=1.2in}
\psfig{figure=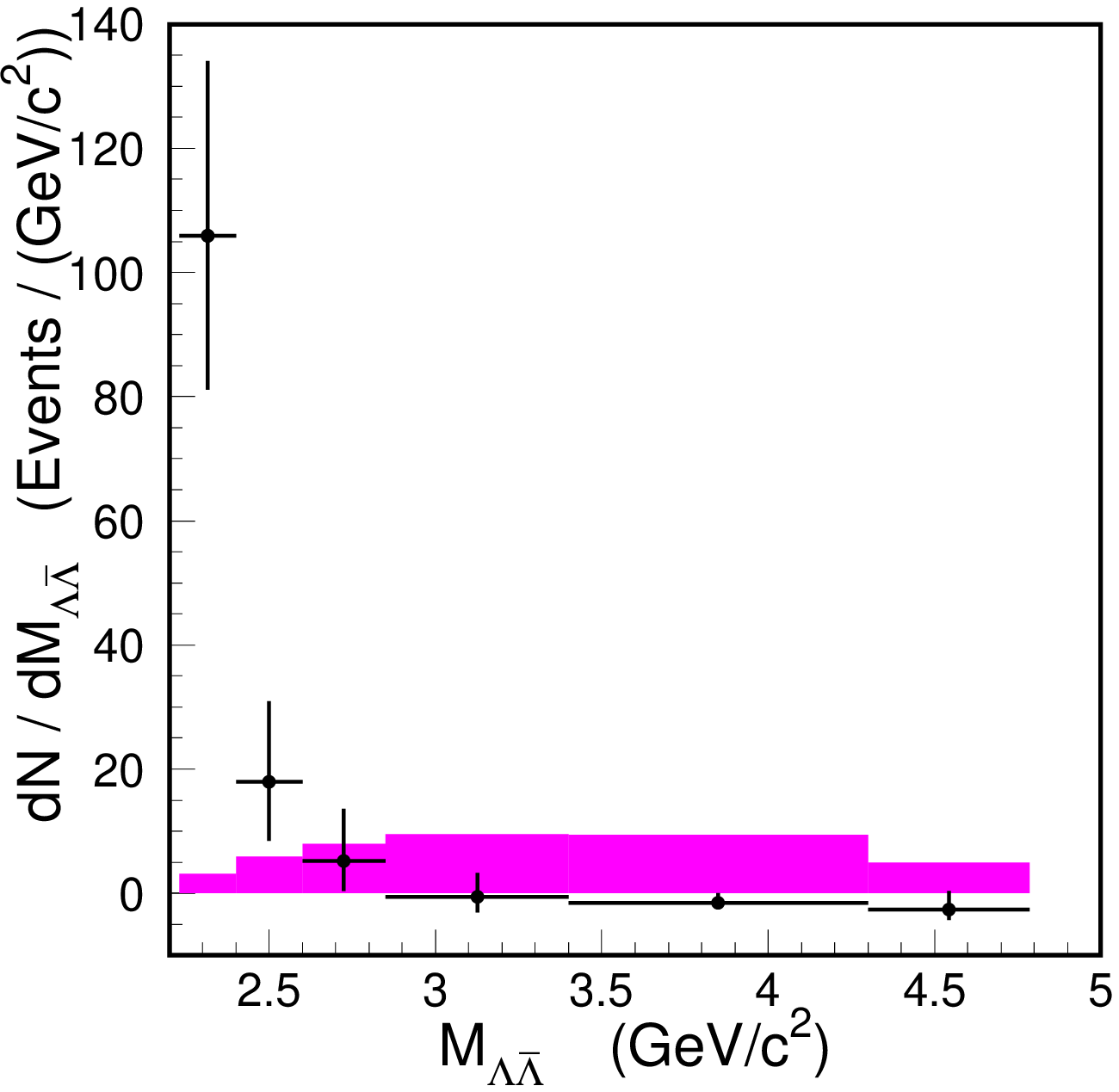,height=1.2in}
\psfig{figure=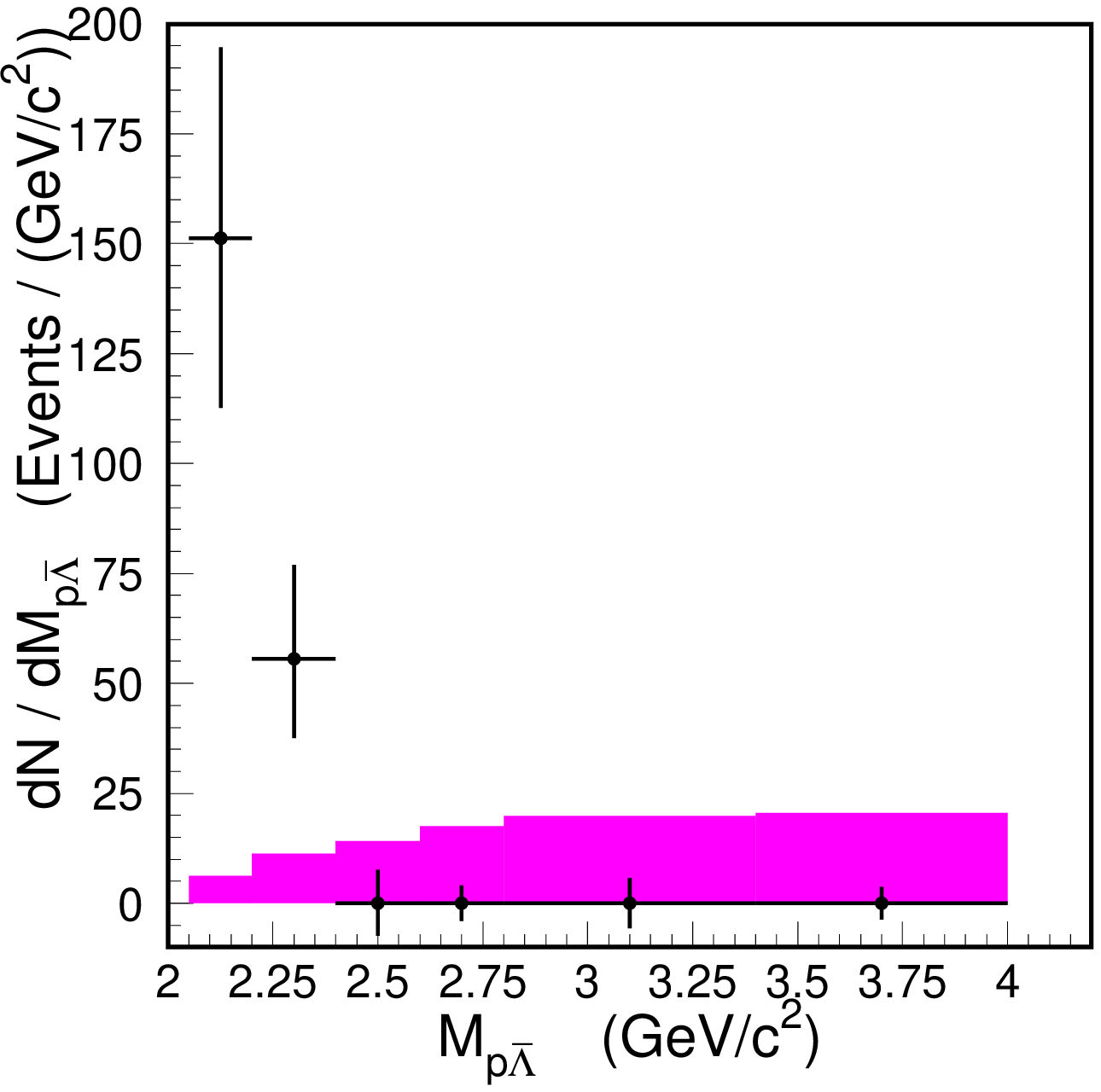,height=1.2in}
\put(-390,62){{\sf\shortstack[c]{\footnotesize (a)}}}
\put(-300,62){{\sf\shortstack[c]{\footnotesize (b)}}}
\put(-209,62){{\sf\shortstack[c]{\footnotesize (c)}}}
\put(-119,62){{\sf\shortstack[c]{\footnotesize (d)}}}
\put(-28,62){{\sf\shortstack[c]{\footnotesize (e)}}}
\caption{Differential branching fractions for (a) $\ppkp$, (b) $\ppks$, (c)
\plpi, (d) \llk{} 
and (e) \plg{} as a function of the baryon pair mass. 
The shaded distribution shows the expectation
from phase-space simulations.
\label{fig:bfmass}}
\end{figure}

\begin{figure}
\vskip 0.2cm
\psfig{figure=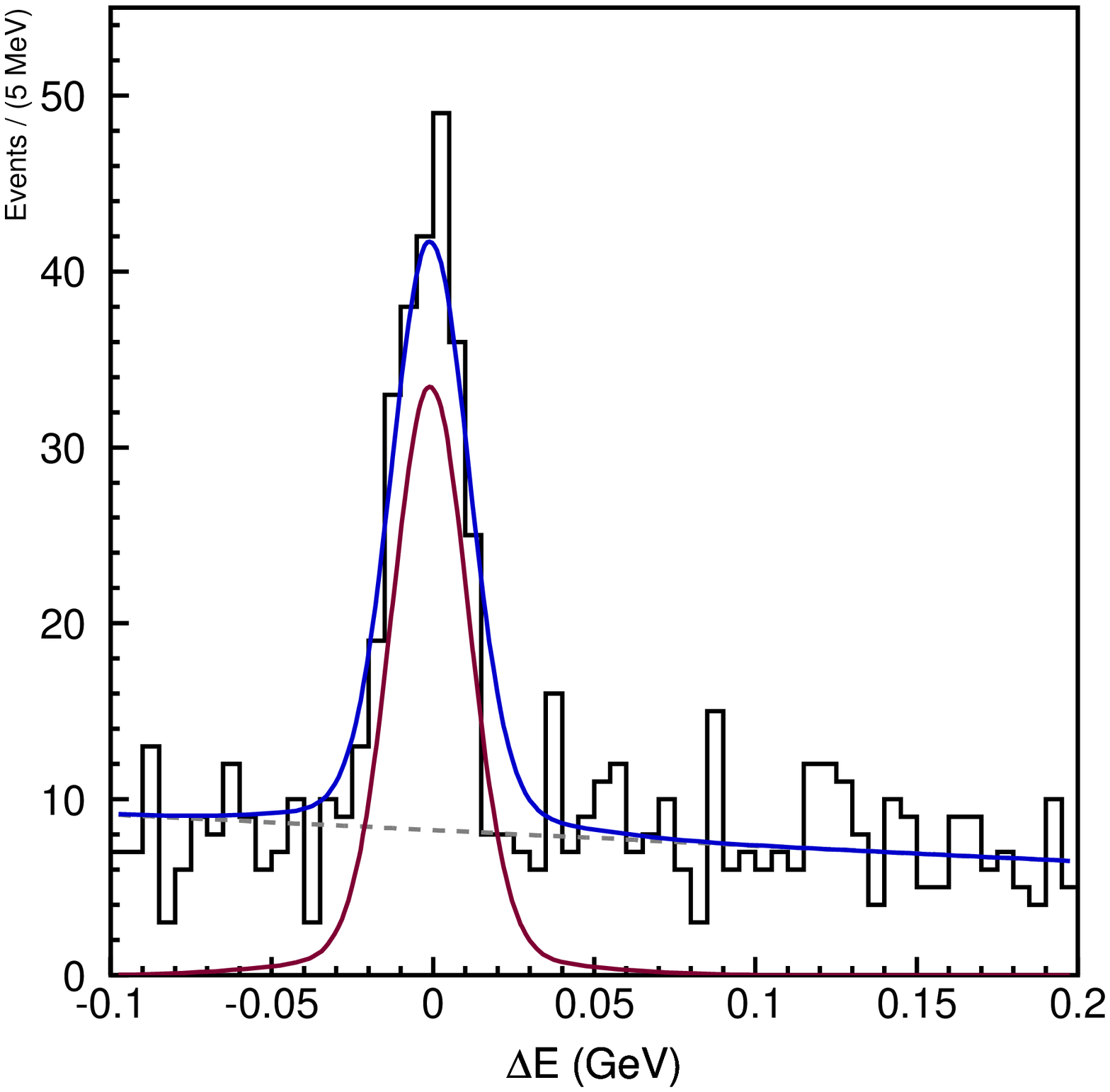,height=1.2in}
\psfig{figure=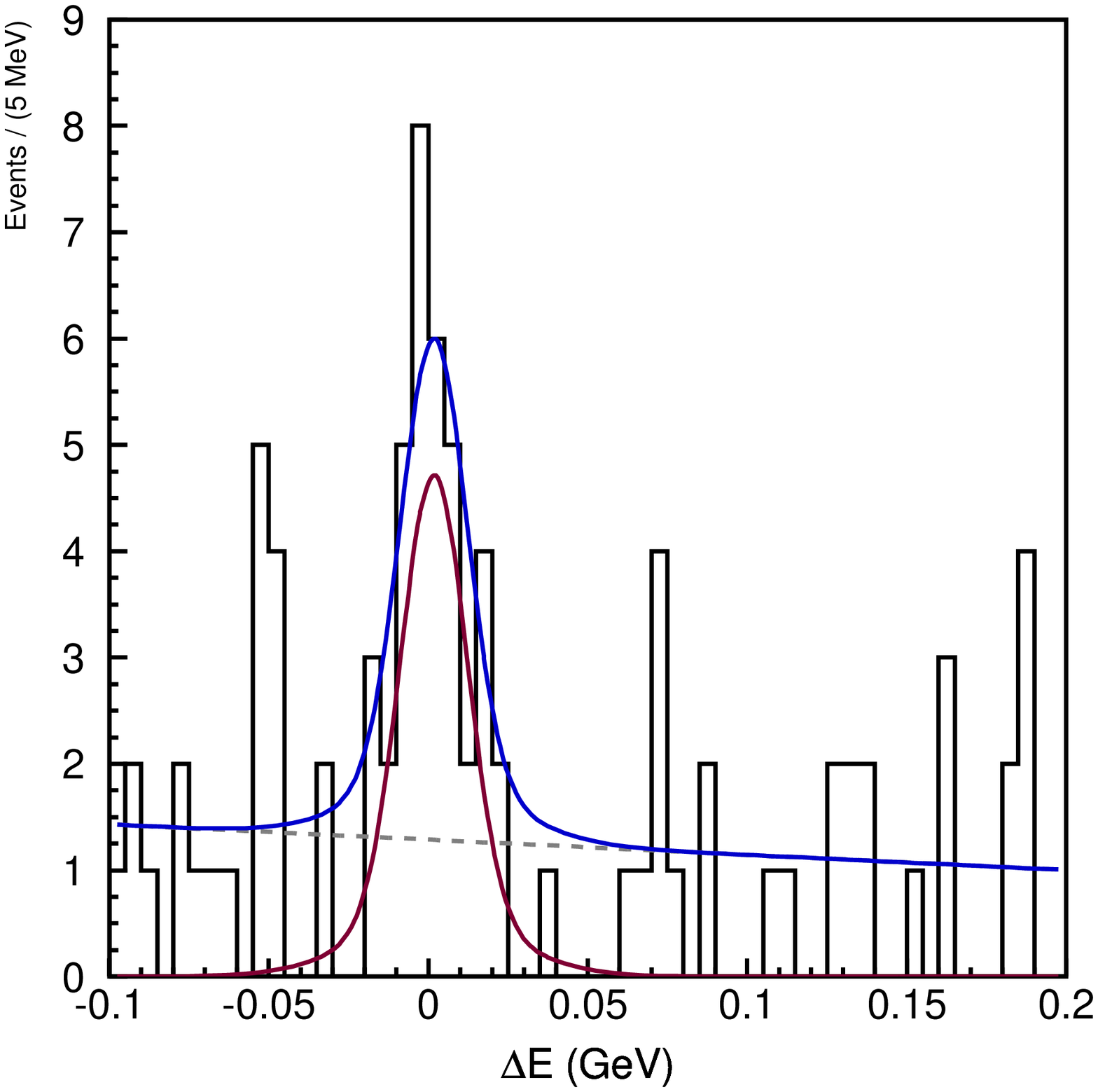,height=1.2in}
\psfig{figure=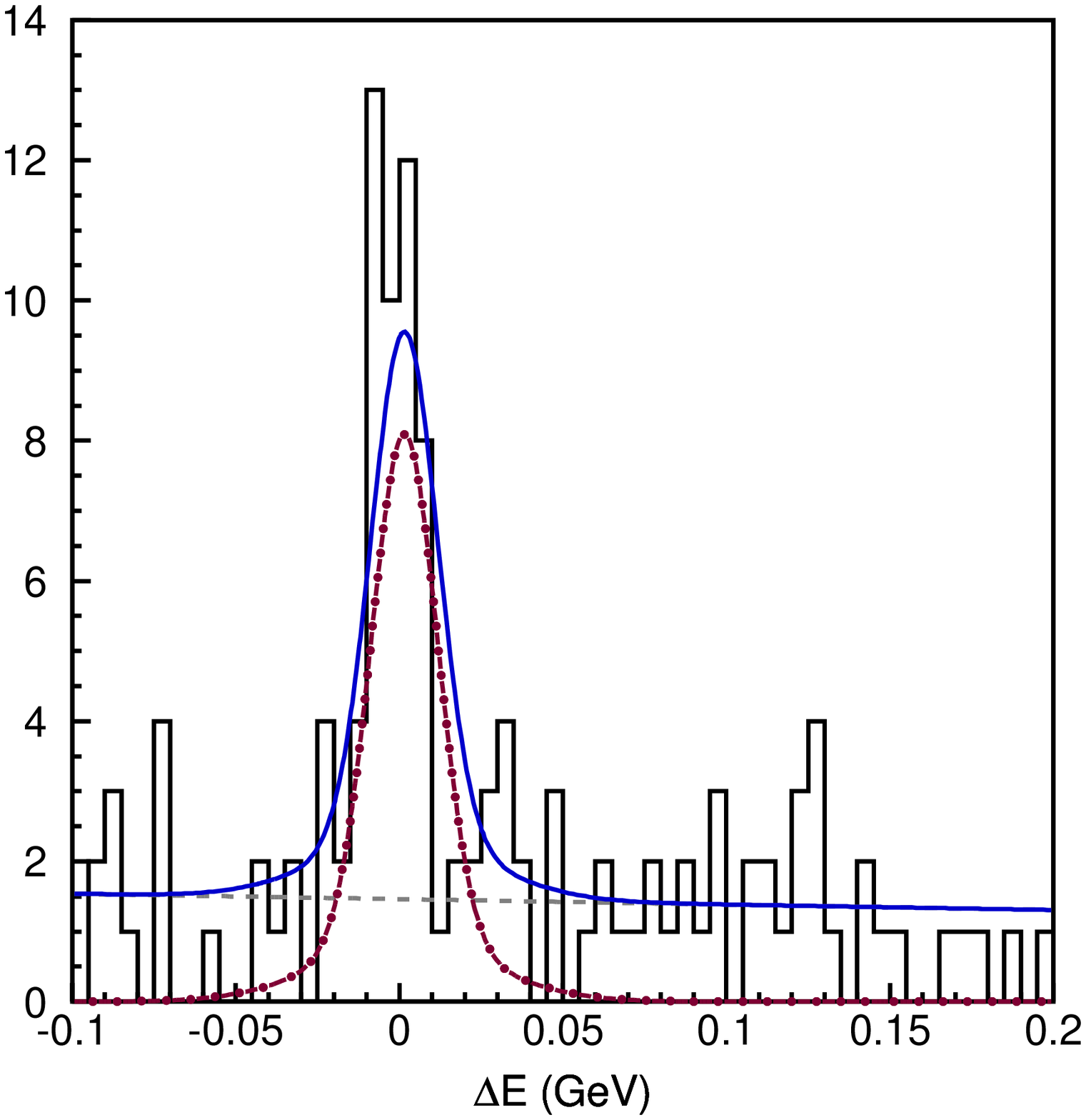,height=1.2in}
\psfig{figure=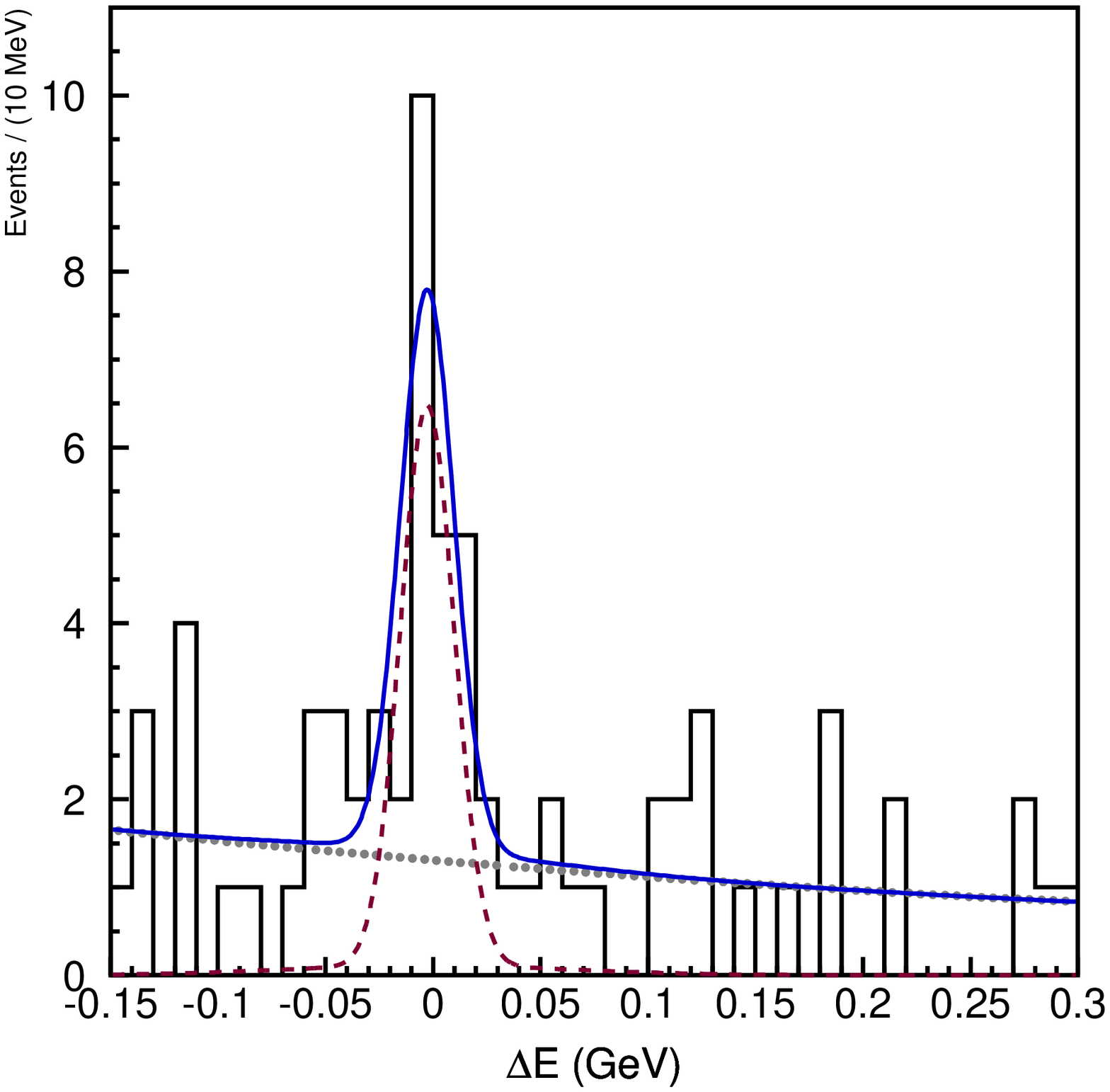,height=1.2in}
\psfig{figure=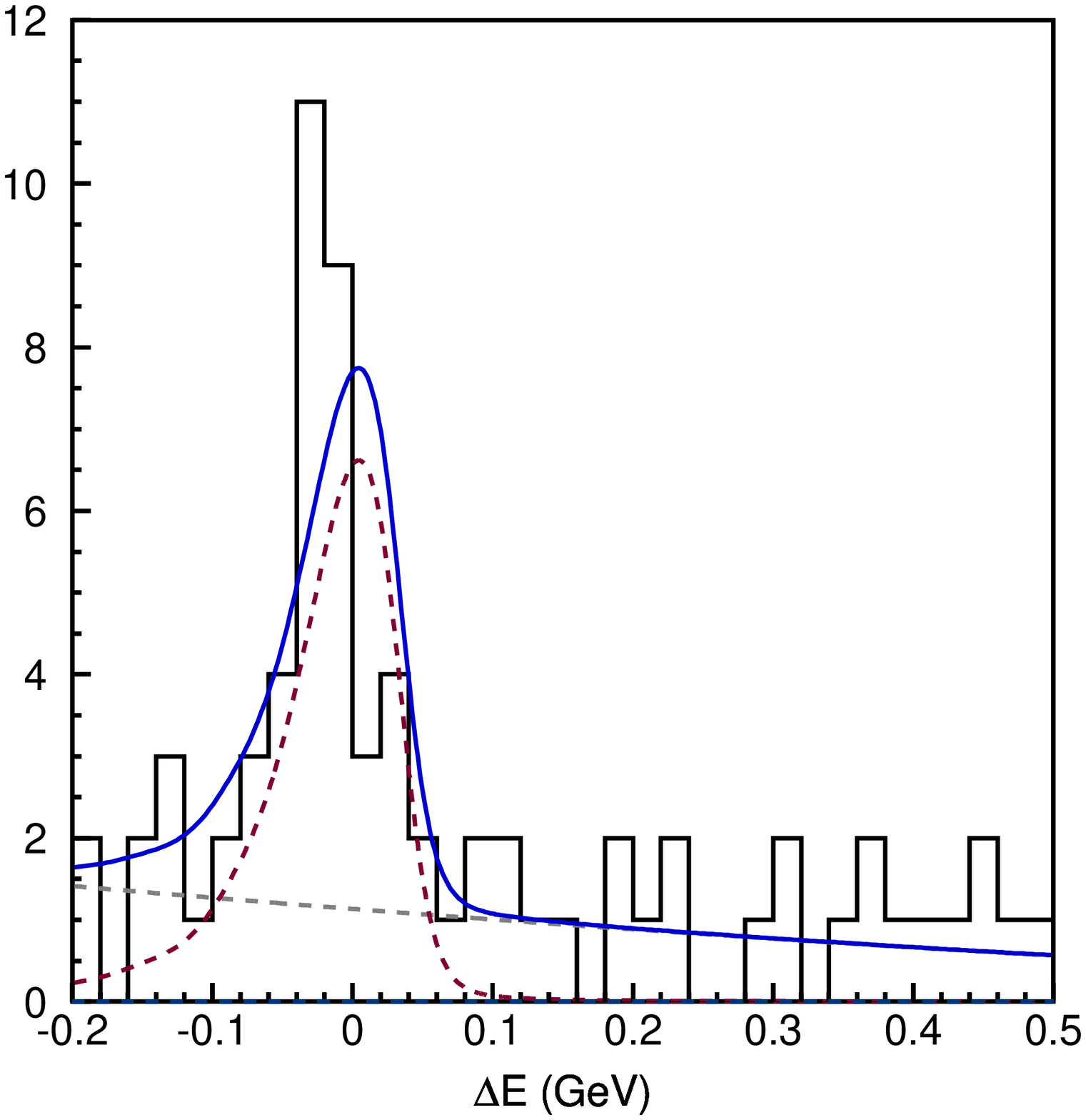,height=1.2in}
\put(-386,65){{\sf\shortstack[c]{\footnotesize (a)}}}
\put(-296,65){{\sf\shortstack[c]{\footnotesize (b)}}}
\put(-205,65){{\sf\shortstack[c]{\footnotesize (c)}}}
\put(-115,65){{\sf\shortstack[c]{\footnotesize (d)}}}
\put(-24,65){{\sf\shortstack[c]{\footnotesize (e)}}}
\caption{Distributions of the projections in \de{} for
(a) $\ppkp$, (b) $\ppks$, (c)
\plpi, (d) \llk{} and (e) \plg{} with baryon-antibaryon mass less than 2.85
GeV/$c^2$. The solid, dotted and dashed lines represent the combined fit
results, fitted signal and fitted background, respectively.
\label{fig:bfde}}
\end{figure}

The differential branching fractions as a function of the baryon pair mass are
shown in Figure~\ref{fig:bfmass} after applying a charmonium veto for the \ppk{}
modes. A clear enhancement at threshold in disagreement with phase space
expectations can be seen. The width of the peaking behaviour depends on the
signal mode.
The \de{} distributions (with $\mb > 5.27$ GeV) for the five three-body baryonic
$B$ decays are shown in Figure~\ref{fig:bfde}. The projections of the fit
results are displayed as solid curves. The results are summarized in
Table~\ref{tab:BF}.

We study the angular distribution of the baryon-antibaryon system in its
helicity frame. The angle $\theta_p$ is defined as the
angle between the proton direction and the meson direction in the 
baryon-antibaryon pair rest frame.  
We find an asymmetry in the angular distributions, 
which indicates that the fragmentation picture is favored. 
Antiprotons are emitted along
the $K^+$ direction most of the time, which can be explained by 
a parent $\bar{b} \to \bar{s}$ penguin transition followed by
$\bar{s} u$ fragmentation into the final state. 
The energetic $\bar{s}$ quark picks up the $u$ quark from a
$u\bar{u}$ pair in vacuum and the remaining $\bar{u}$ quark then
drags a $\bar{u}\bar{d}$ diquark out of vacuum. 
%

Two-body charmless baryonic $B$ decays were investigated, but no
signal events were seen. For a detailed description of this analysis please
read~\cite{bib:pp}. The upper limits are given in Table~\ref{tab:BF}.

The newly observed narrow pentaquark state, $\Theta^+$~\cite{penta}, can
decay into $p \ks$. We perform a search in $B^0 \to \sppk_S$
by requiring   $1.53$ GeV/$c^2 < M_{p\ks} < 1.55 $ GeV/$c^2$.
We find no evidence for a pentaquark signal.
We also perform a search for $\Theta^{++}$, 
which can decay to $p K^+$ in the mode $B^+ \to \sppk^+$~\cite{pentappk}.
We find no evidence for signal. We set an upper limit assuming this state is 
narrow and centered near 1.71 GeV/$c^2$. The upper limits are listed in
Table~\ref{tab:BF}.

Systematic errors are studied; for a detailed description of the procedure
please refer to~\cite{bib:lamc,bib:pp,bib:lowm,LLK}. 
The final systematic errors are
listed in Table~\ref{tab:BF}

\subsection{\lcppi}

\begin{figure}[b]
\vskip 0.2cm
\psfig{figure=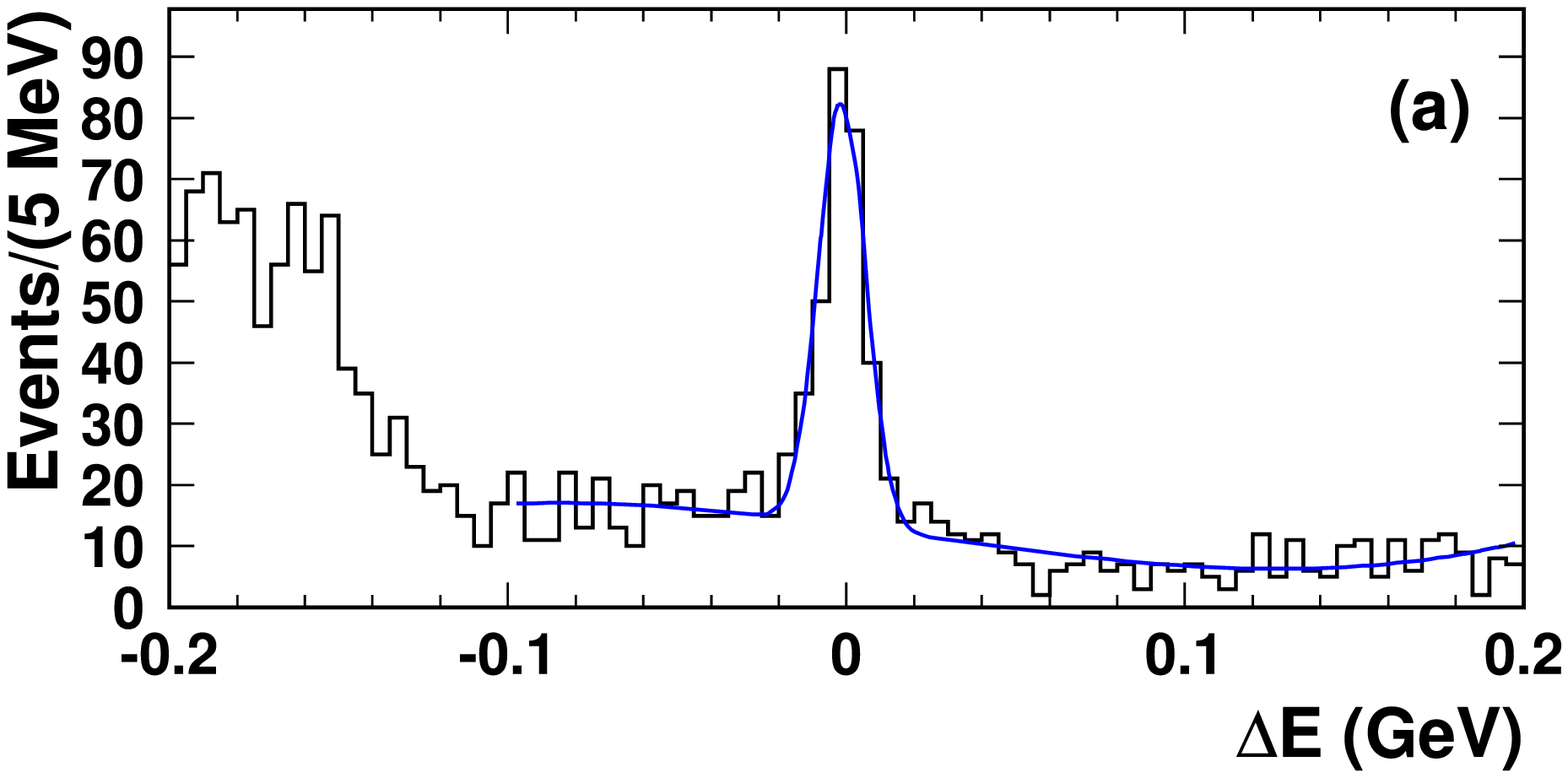,height=0.725in}
\psfig{figure=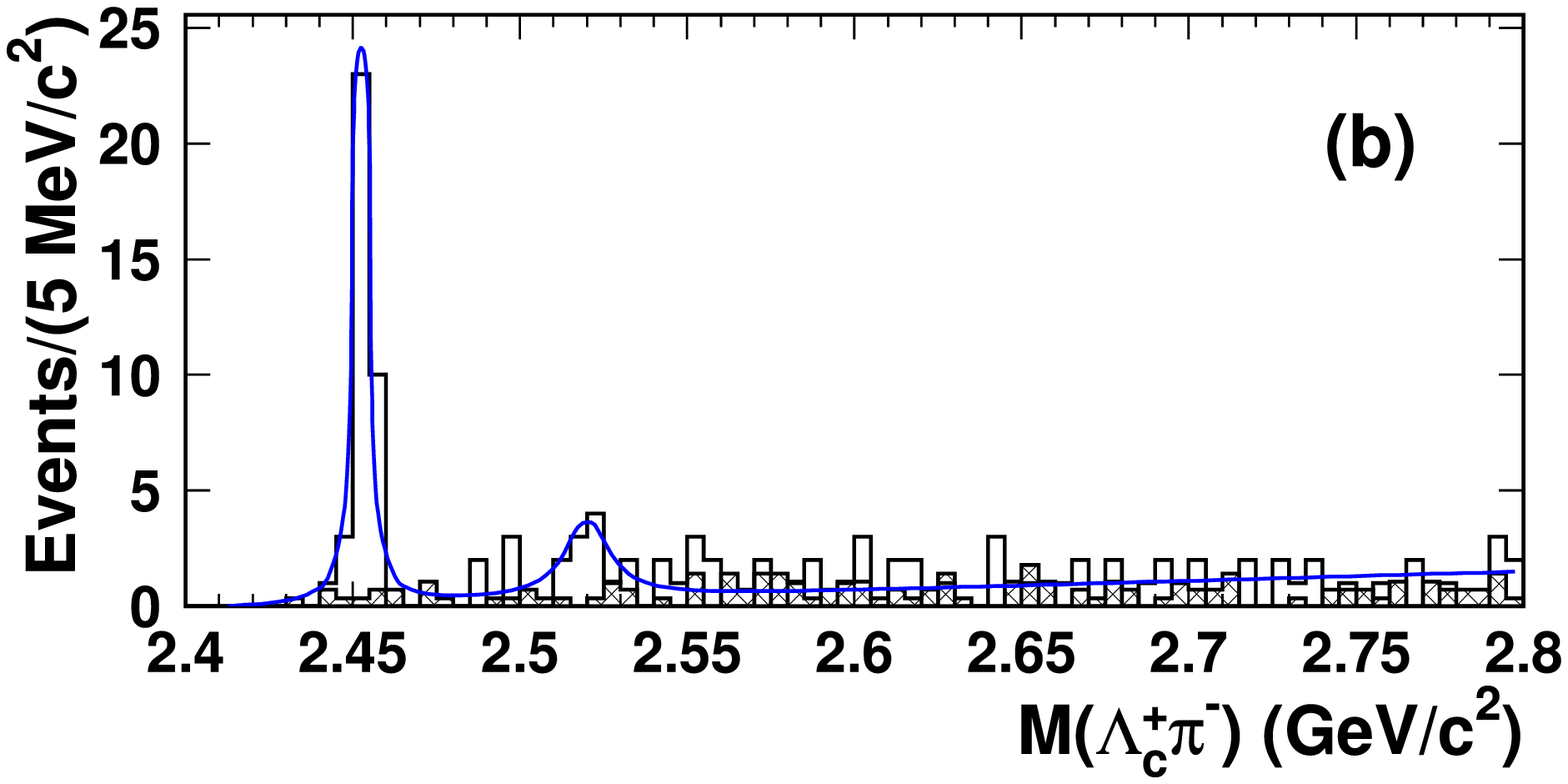,height=0.725in}
\psfig{figure=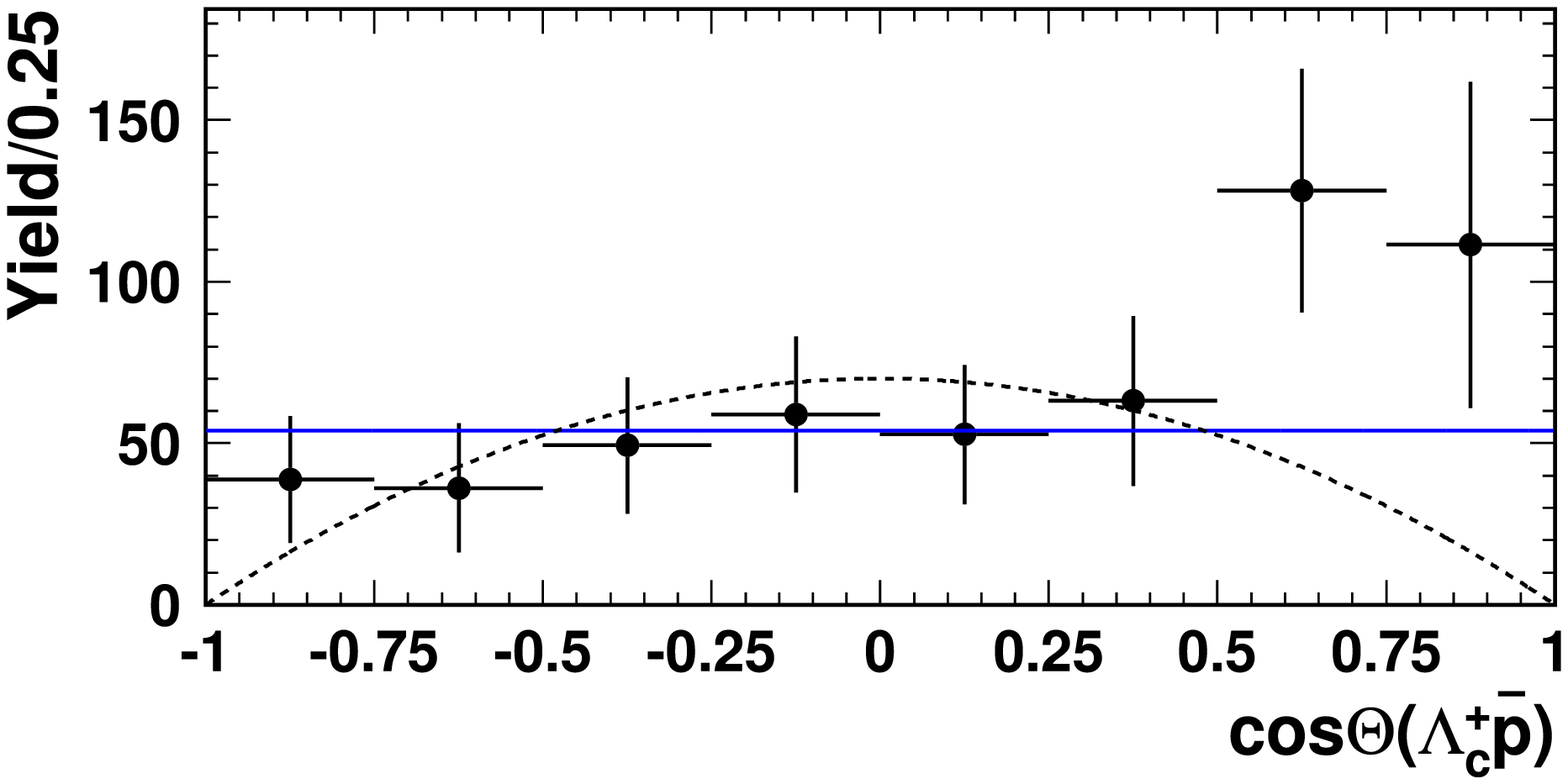,height=0.725in}
\psfig{figure=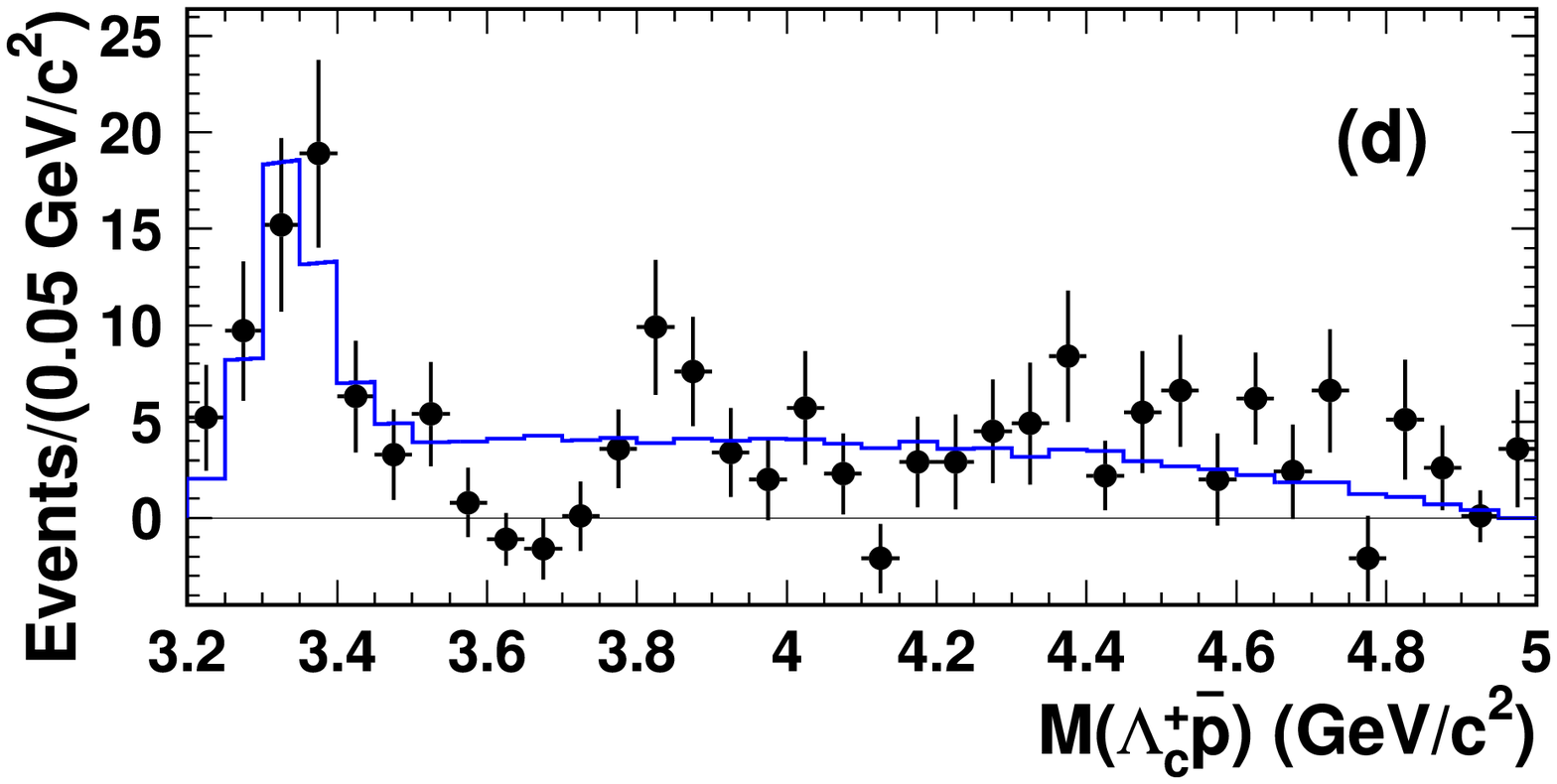,height=0.725in}
\put(-122,43){{\sf\shortstack[c]{\tiny (c)}}}
\caption{\de, $\Lambda_c^+ \pi^-$ mass, $\Lambda_c^+ \bar{p}$ helicity 
and its mass distribution.
\label{fig:lamc}}
\end{figure}
We perform a Dalitz plots analysis to investigate of the three-body charmed 
\lcppi{} decay. The distributions for \de, the mass of the $\Lambda_c^+\pi^-$
pair, the mass of the $\Lambda_c^+\bar{p}$ pair and its helicity distribution
are displayed in Figure~\ref{fig:lamc}. The intermediate states of
$\Sigma_c(2455)^0$ and $\Sigma_c(2520)^0$ can clearly be identified in
Figure~\ref{fig:lamc} (b). The branching fractions of these two-body
intermediate states are given in Table~\ref{tab:BF}.
As in the previous analyses, the baryon-antibaryon
pair shows a low mass enhancement. This can be parameterized with a Breit-Wigner
peak and feed downs. The fit gives a mass of $(3.35^{+0.01}_{-0.02})$ GeV/$c^2$
and a full width of $(0.07^{+0.04}_{-0.03})$ GeV/$c^2$. The fit yield is
$50\pm 10$ events with a statistical significance of $5.6\sigma$. The angular
distribution as seen in Figure~\ref{fig:lamc} (c) cannot conclusively determine
whether the enhancement arises from a resonance, fragmentation or final state
interactions.

\section{Summary}

We have observed three-body charmless baryonic $B$ decays and charmed
baryonic two- and three-body decays. All three-body decays display an enhancement
at threshold in the baryon-antibaryon mass. The fragmentation picture seems to
be favored to explain this behaviour. The results are listed in
Table~\ref{tab:BF}.

\begin{table}[t]
\caption{Branching fractions of the baryonic $B$ decays. We list the decay
modes, the significances, the branching fractions and the experiment that
performed the analysis.
\label{tab:BF}}
\vspace{0.3cm}
\begin{center}
\begin{tabular}{|c|c|c|c|}
\hline
& & & \\
Mode	&  Significance & Branching fraction & Experiment \\
	& $\sigma$	&  $[10^{-6}]$		&	\\
\hline
\BF (\llk ) & 7.4	& $2.91\,^{+0.90}_{-0.70}\pm 0.38$ & Belle	\\
\hline
\BF (\llpi )	& -	& $<2.8$		& Belle \\
\hline
\BF ($\ppkp$) & $>10$ & $5.30\,^{+0.45}_{-0.39}\pm 0.58$ & Belle \\
		   & $>10$ & $6.7\pm 0.9 \pm 0.6$ & BaBar \\
\hline
\BF ($\ppks$) & $>10$		& $1.20\,^{+0.32}_{-0..22}\pm 0.14$ & Belle \\
\hline
\BF (\plpi ) & 11.2		& $3.27\,^{+0.62}_{-0.51}\pm 0.39$ & Belle \\
\hline
\BF (\plg ) & 8.6		& $2.16\,^{+0.58}_{-0.53}\pm 0.20$ & Belle \\
\hline
\BF $(p\bar{\Sigma}^0\gamma)$ & - & $<4.6$			& Belle \\
\hline
\BF (\lcppi ) & 18.1		& $201\pm 15 \pm 20 \pm 52$	& Belle \\
\BF $(\Lambda_c^+\bar{p}$ structure) & 6.2 & $38.7\,^{+7.7}_{-7.2}\pm 4.3 \pm
10.1$ & Belle \\
\BF $(\Sigma_c(2455)\bar{p})$	& 8.4	& $36.7\,^{+7.4}_{-6.6}\pm 3.6 \pm 9.5$ &
Belle \\
\hline
\BF $(B\to p\bar{p}/\Lambda\bar{p}/\Lambda\bar{\Lambda})$	
	& - 	& $<0.41 / 0.49 / 0.69$		& Belle \\
\hline
\BF $(B\to \theta^+p)\times \BF (\theta^+ \to K_S^0 p)$ 
		& -		& $<0.23$	& Belle \\
\hline
\BF $(B\to \theta^{++}p)\times \BF (\theta^{++}K^+p)$ 
		& - 		& $<0.091$ / $<0.15-0.40$ & Belle / BaBar \\
\hline
\end{tabular}
\end{center}
\end{table}

\end{document}